\documentclass[preprintnumbers,amsmath,amssymbm,prd]{revtex4}
\usepackage{epsfig}
\usepackage{graphicx}
\usepackage{amssymb}

\begin{document}
\title{Stationary scalar clouds supported by rapidly-rotating acoustic black holes in a photon-fluid model}
\author{Shahar Hod}
\affiliation{The Ruppin Academic Center, Emeq Hefer 40250, Israel}
\affiliation{ } \affiliation{The Hadassah Institute, Jerusalem
91010, Israel}
\date{\today}

\begin{abstract}
\ \ \ It has recently been proved that, in the presence of vortex flows, 
the fluctuation dynamics of a rotating photon-fluid model is governed by the Klein-Gordon equation of an effective massive
scalar field in a $(2+1)$-dimensional acoustic black-hole spacetime. 
Interestingly, it has been demonstrated numerically that the rotating acoustic black hole, like the familiar 
Kerr black-hole spacetime, may support spatially regular stationary density fluctuations (linearized acoustic scalar `clouds') 
in its exterior regions. In particular, it has been shown that the composed 
rotating-acoustic-black-hole-stationary-scalar-field configurations of 
the photon-fluid model exist in the narrow dimensionless 
regime $\alpha\equiv\Omega_0/m\Omega_{\text{H}}\in(1,\alpha_{\text{max}})$ with $\alpha_{\text{max}}\simeq 1.08$ 
[here $\Omega_{\text{H}}$ is the angular velocity of the black-hole horizon and $\{\Omega_0,m\}$ 
are respectively the effective proper mass and the azimuthal harmonic index of the acoustic scalar field]. 
In the present paper we use analytical techniques in order to explore the physical and mathematical properties of
the acoustic scalar clouds of the photon-fluid model in the regime $\Omega_{\text{H}}r_{\text{H}}\gg1$ 
of rapidly-spinning central supporting acoustic black holes. 
In particular, we derive a remarkably compact analytical formula for the 
discrete resonance spectrum $\{\Omega_0(\Omega_{\text{H}},m;n)\}$ which 
characterizes the stationary bound-state acoustic scalar clouds of the photon-fluid model. 
Interestingly, it is proved that the critical (maximal) mass
parameter $\alpha_{\text{max}}$, which determines the regime of existence of the 
composed acoustic-black-hole-stationary-bound-state-massive-scalar-field configurations, 
is given by the exact dimensionless relation $\alpha_{\text{max}}=\sqrt{{{32}\over{27}}}$. 

\end{abstract}
\bigskip
\maketitle


\section{Introduction}

Kerr black-hole spacetimes \cite{Kerr} are characterized by the 
presence of an ergoregion \cite{Chan}, a region outside the black-hole horizon in which matter fields 
are bound to co-rotate with the central spinning black holes. 
Interestingly, it has been demonstrated, both analytically \cite{Hodrc} and numerically \cite{HerR},
that this physically intriguing feature of spinning black-hole spacetimes allows them to support 
spatially regular stationary configurations of bosonic (integer-spin) fields that co-rotate with the central
black hole. 

The stationary hairy scalar field configurations, which in the linearized regime have received the nickname 
`scalar clouds' in the physics literature \cite{Hodrc,HerR}, are characterized by proper frequencies that are 
in resonance with the angular velocity $\Omega_{\text{H}}$ of the 
central supporting spinning black hole \cite{Hodrc,HerR}. In particular, the characteristic proper 
frequency of a stationary scalar cloud with an azimuthal harmonic index $m$ 
coincides with the critical (marginal) frequency for the superradiant scattering phenomenon \cite{Zel,PressTeusup}
in the rotating black-hole spacetime \cite{Hodrc,HerR,Noteunits}:
\begin{equation}\label{Eq1}
\omega=m\Omega_{\text{H}}\ .
\end{equation}

In addition, spatially regular (bounded) bosonic clouds are characterized by the simple 
upper bound \cite{Hodrc,HerR}
\begin{equation}\label{Eq2}
\omega^2<\mu^2\  ,
\end{equation}
where $\mu$ is the proper mass \cite{Notedim} of the supported stationary scalar field.  
The relation (\ref{Eq2}) implies that the co-rotating massive scalar field configurations are spatially bounded to the central 
black hole and cannot radiate their energy and angular momentum to infinity.

Intriguingly, an analogous physical phenomenon in a rotating photon-fluid system 
has recently been revealed in the highly important work \cite{Cim}. 
Photon-fluids are nonlinear optical systems whose physical and mathematical properties can
be described by the hydrodynamic equations of an interacting Bose gas \cite{Cim,PF1,PF2,Un}. 
In particular, it has been shown \cite{Cim,Mar1} 
that photon-fluid systems are 
characterized by the presence of long-wavelength elementary excitations (phonons) that
behave as massive scalar fields in an effective acoustic curved spacetime. 

The dynamics of massive phonons over a draining vortex flow in the photon-fluid model has been 
investigated recently in \cite{Cim} as the acoustic analogue of the (more familiar) dynamics of 
massive scalar fields in rotating curved black-hole spacetimes. 
In particular, it has been explicitly proved \cite{Cim} that the dynamics of linearized acoustic excitations 
in the photon-fluid model with a draining vortex flow are governed by the familiar Klein-Gordon equation of an effective 
scalar field of proper mass $\Omega_0$ that propagates in an acoustic $(2+1)$-dimensional spinning 
black-hole spacetime which, like the familiar Kerr black-hole spacetime, possesses an ergoregion. 

Intriguingly, using direct numerical techniques, it has been explicitly demonstrated in \cite{Cim} that 
the acoustic spinning black-hole spacetime may support stationary linearized 
density fluctuations (acoustic scalar `clouds') in its exterior regions. In particular, it has been revealed \cite{Cim}
that the composed acoustic-black-hole-stationary-massive-scalar-field configurations of 
the photon-fluid model are characterized by the narrow regime of existence  \cite{Cim,Notemmm}
\begin{equation}\label{Eq3}
\alpha\equiv{{\Omega_0}\over{m\Omega_{\text{H}}}}\in(1,\alpha_{\text{max}})\ \ \ \ \text{with}\ \ \ \ 
\alpha_{\text{max}}\simeq 1.08\  ,
\end{equation}
where $\Omega_{\text{H}}$ is the angular velocity that characterizes the acoustic 
horizon of the central supporting spinning black hole. 

The main goal of the present paper is to explore, using analytical techniques, the physical and mathematical properties 
of the composed acoustic-spinning-black-hole-stationary-linearized-scalar-field configurations of 
the photon-fluid model. In particular, we shall derive a remarkably compact analytical formula for the 
discrete resonance spectrum $\{\Omega_0(\Omega_{\text{H}},m;n)\}$ \cite{Notennnn} that
characterizes the spatially regular stationary acoustic scalar clouds 
in the dimensionless regime $\Omega_{\text{H}}r_{\text{H}}\gg1$ \cite{Noterhrh} of rapidly-spinning central supporting black holes. 
In addition, we shall provide a simple analytical explanation for the existence of the numerically observed \cite{Cim} interesting 
upper bound $\alpha<\alpha_{\text{max}}\simeq1.08$ [see Eq. (\ref{Eq3})] on the regime of existence of the composed acoustic-black-hole-stationary-bound-state-massive-scalar-field configurations.

\section{Description of the system}

We study the physical and mathematical properties of density fluctuations in a rotating photon-fluid model. 
Intriguingly, a formal equivalence has recently been established in the physically important work \cite{Cim} 
between the dynamics of linearized acoustic phonons that propagate
on top of an inhomogeneous photon-fluid and the dynamics of linearized massive scalar fields in 
a spinning curved spacetime. In particular, it has been explicitly proved in \cite{Cim} that, in the presence of vortex flows, 
the dynamics of acoustic density fluctuations in 
the long-wavelength regime of the photon-fluid model are governed by the
Klein-Gordon equation of a massive scalar field in an effective $(2+1)$-dimensional curved spacetime. 

The effective acoustic spacetime of the $(2+1)$-dimensional rotating photon-fluid model 
can be described, using polar coordinates, by the non-trivial curved line element \cite{Cim,Notert}
\begin{equation}\label{Eq4}
ds^2=-\Big(1-{{r_{\text{H}}}\over{r}}-{{\Omega^2_{\text{H}}r^4_{\text{H}}}\over{r^2}}\Big)dt^2+
\Big(1-{{r_{\text{H}}}\over{r}}\Big)^{-1}dr^2-
2\Omega_{\text{H}}r^2_{\text{H}}d\theta dt+r^2 d\theta^2\  .
\end{equation}
Here $r_{\text{H}}$ is the radius of the acoustic black-hole horizon, which is determined 
as the circular ring at which the 
inward radial velocity $v_{\text{r}}$ of the fluid flow equals the speed of sound $c_{\text{s}}$ \cite{Cim,Notegrr,Notecs}. 
The physical parameter $\Omega_{\text{H}}$ in the curved line element (\ref{Eq4}) is the angular velocity of the 
effective acoustic horizon. 

Interestingly, like the familiar Kerr black-hole solution of the Einstein field equations, the rotating acoustic 
spacetime (\ref{Eq4}) of the photon-fluid model 
is characterized by the presence of an effective ergoregion whose outer boundary \cite{Cim}
\begin{equation}\label{Eq5}
r_{\text{E}}={{1}\over{2}}r_{\text{H}}\Big(1+\sqrt{1+4\Omega^2_{\text{H}}r^2_{\text{H}}}\Big)\ 
\end{equation}
is determined by the condition $g_{tt}=0$. 

As explicitly proved in \cite{Cim}, the spatial behavior of density fluctuations of the form
\begin{equation}\label{Eq6}
\rho(t,r,\theta)={{\psi(r)}\over{\sqrt{r}}}e^{im\theta-i\Omega t}\
\end{equation}
in the effective acoustic spacetime (\ref{Eq4}) of the rotating photon-fluid model are governed by the radial 
differential equation 
\begin{equation}\label{Eq7}
\Big[\Delta{{d}\over{dr}}\Big(\Delta{{d}\over{dr}}\Big)-V(r;\Omega)\Big]\psi(r)=0\  ,
\end{equation}
where 
\begin{equation}\label{Eq8}
\Delta\equiv 1-{{r_{\text{H}}}\over{r}}\  .
\end{equation}
The effective radial potential of the photon-fluid system is given by the functional expression
\begin{equation}\label{Eq9}
V(r;\Omega)=-\Big(\Omega-{{m\Omega_{\text{H}}r^2_{\text{H}}}\over{r^2}}\Big)^2+
\Delta\Big(\Omega^2_0+{{m^2}\over{r^2}}+{{r_{\text{H}}}\over{2r^3}}-{{\Delta}\over{4r^2}}\Big)\  .
\end{equation}
The physical parameter $\Omega_0$, which plays the role of an effective scalar mass, is the rest
energy of the collective excitations (phonons) \cite{Cim}. The $\theta$-periodicity of the angular function $e^{im\theta}$ 
in the field decomposition (\ref{Eq6}) implies that the azimuthal harmonic index $|m|$ of the scalar perturbation modes 
is an integer \cite{Notem11}.

In the next section we shall use analytical techniques in order to derive the discrete resonance spectrum 
$\{\Omega_0(\Omega_{\text{H}},m;n)\}^{n=\infty}_{n=0}$ of the composed acoustic-black-hole-stationary-bound-state-linearized-massive-scalar-field configurations. 
The radial eigenfunctions that characterize the stationary scalar clouds of the acoustic curved spacetime (\ref{Eq4})
are determined by the 
ordinary differential equation (\ref{Eq7}) with the physically motivated boundary conditions of 
spatially regular (bounded) scalar eigenfunctions at the acoustic black-hole horizon and at 
spatial infinity \cite{Cim}:
\begin{equation}\label{Eq10}
\psi(r=r_{\text{H}})<\infty\
\end{equation}
and
\begin{equation}\label{Eq11}
\psi(r\to\infty)\sim e^{-{\sqrt{\Omega^2_0-\Omega^2}r}}\ \ \ \ \text{for}\ \ \ \ \Omega^2<\Omega^2_0\  .
\end{equation}

\section{The discrete resonance spectrum of the composed acoustic-black-hole-scalar-clouds configurations of the photon-fluid model}

In the present section we shall analyze the discrete resonance spectrum $\{\Omega_0(\Omega_{\text{H}},m;n)\}$  that 
characterizes the composed acoustic-black-hole-stationary-bound-state-linearized-massive-scalar-field configurations 
of the photon-fluid model \cite{Cim}. 
The stationary bound-state scalar clouds of the effective rotating black-hole spacetime (\ref{Eq4}) are characterized 
by the resonance condition \cite{Cim}
\begin{equation}\label{Eq12}
\Omega=m\Omega_{\text{H}}<\Omega_0\  .  
\end{equation}

Interestingly, we shall now prove that the resonance spectrum $\{\Omega_0(\Omega_{\text{H}},m;n)\}$ 
of the acoustic scalar clouds can be studied {\it analytucally} in the eikonal large-frequency regime \cite{Notenub}
\begin{equation}\label{Eq13}
\Omega_{\text{H}}r_{\text{H}}\gg m\  
\end{equation}
of the central supporting spinning black hole. 

To this end, it is convenient to write the radial differential equation (\ref{Eq7}), which determines the 
spatial behavior of the scalar eigenfunctions in the acoustic black-hole spacetime (\ref{Eq4}), in the form of 
the mathematically compact Schr\"odinger-like ordinary differential equation 
\begin{equation}\label{Eq14}
{{d^2\psi}\over{dy^2}}-V(y)\psi=0\  ,
\end{equation}
where the tortoise radial coordinate $y$ is defined by the differential relation \cite{Notemap} 
\begin{equation}\label{Eq15}
dy={\Delta}^{-1}dr\  .
\end{equation}

Substituting into Eq. (\ref{Eq9}) the resonant frequency $\Omega=m\Omega_{\text{H}}$ 
[see Eq. (\ref{Eq1})], which characterizes the stationary acoustic scalar clouds of the 
photon-fluid model, one obtains the functional expression
\begin{equation}\label{Eq16}
V(r)=-(m\Omega_{\text{H}})^2\cdot\Big(1-{{r^2_{\text{H}}}\over{r^2}}\Big)^2+
\Delta\Big(\Omega^2_0+{{m^2}\over{r^2}}+{{r_{\text{H}}}\over{2r^3}}-{{\Delta}\over{4r^2}}\Big)\
\end{equation}
for the effective radial potential $V[r(y)]$ of the 
composed acoustic-black-hole-stationary-bound-state-massive-scalar-field configurations. 

We shall now show explicitly that the Schr\"odinger-like ordinary differential equation (\ref{Eq14}) 
for the spatially regular stationary scalar clouds in the rotating acoustic black-hole spacetime (\ref{Eq4}) 
is amenable to a standard WKB analysis \cite{WKB1,WKB2,WKB3,Will,Hodalm} in the dimensionless regime 
$\Omega_{\text{H}}r_{\text{H}}\gg1$ of rapidly spinning supporting black holes.

Interestingly, the potential (\ref{Eq16}) of the composed 
acoustic-black-hole-stationary-scalar-clouds configurations has an effective binding well in the vicinity of 
the acoustic horizon. 
As shown in \cite{WKB1,WKB2,Hodalm}, in the eikonal large-frequency regime (\ref{Eq13}), 
one can express the WKB resonance condition for 
the bound-state field configurations of the one-dimensional Schr\"odinger-like ordinary differential equation (\ref{Eq14}) 
in the remarkably compact form
\begin{equation}\label{Eq17}
{V_{\text{min}}\over{\sqrt{2V^{''}_{\text{min}}}}}=-(n+{1\over 2})\ \ \ ; \ \ \ n=0,1,2,...\  ,
\end{equation}
where $V^{''}\equiv d^2V/dy^2$. The effective binding potential $V_{\text{min}}$ and 
its second spatial derivative $V^{''}_{\text{min}}$ in the WKB resonance condition (\ref{Eq17}) 
are evaluated at the minimum point $r=r_{\text{min}}$ of the potential (\ref{Eq16}), where 
\begin{equation}\label{Eq18}
V^{'}\equiv {{dV}\over{dy}}=0\ \ \ \ \text{for}\ \ \ \ r(y)=r_{\text{min}}  .
\end{equation}

Substituting the effective binding potential (\ref{Eq16}) of the composed 
spinning-black-hole-acoustic-scalar-clouds configurations into Eq. (\ref{Eq18}), one 
finds the relation
\begin{equation}\label{Eq19}
\Omega^2_0={{4(m\Omega_{\text{H}})^2(r^2_{\text{min}}-r^2_{\text{H}})r_{\text{H}}}\over{r^3_{\text{min}}}}
\cdot\{1+O[(\Omega_{\text{H}}r_{\text{H}})^{-2}]\}\ \ \ \ \text{for}\ \ \ \ \Omega_{\text{H}}r_{\text{H}}\gg1\
\end{equation}
in the eikonal large frequency regime (\ref{Eq13}) of the central supporting acoustic black hole.

Substituting the relation (\ref{Eq19}) into the WKB equation (\ref{Eq17}), one 
finds the resonance equation 
\begin{equation}\label{Eq20}
4r_{\text{H}}(r^2_{\text{min}}-r^2_{\text{H}})-(r_{\text{min}}-r_{\text{H}})(r_{\text{min}}+r_{\text{H}})^2=-{{2r_{\text{H}}\sqrt{2r^2_{\text{min}}-6r^2_{\text{H}}}}\over{m\Omega_{\text{H}}}}\cdot\big(n+{1\over2}\big)\
\end{equation}
for the radial location of the minimum $r=r_{\text{min}}$ of the effective binding potential (\ref{Eq16}) that characterizes 
the composed acoustic-black-hole-stationary-scalar-clouds configurations.

As we shall now show, the (rather cumbersome) resonance equation (\ref{Eq20}) 
can be solved analytically using an iteration scheme. 
The zeroth-order solution $r^{(0)}_{\text{min}}\equiv r_{\text{min}}(\Omega_{\text{H}}r_{\text{H}}\to\infty)$ 
of the resonance equation (\ref{Eq20}) is given by the simple asymptotic value
\begin{equation}\label{Eq21}
r^{(0)}_{\text{min}}=3r_{\text{H}}\  .
\end{equation}
Next, substituting
\begin{equation}\label{Eq22}
r_{\text{min}}=3r_{\text{H}}\cdot\big[1+\alpha(\Omega_{\text{H}}r_{\text{H}})^{-1}\big]\
\end{equation}
into the resonance equation (\ref{Eq20}), one finds
\begin{equation}\label{Eq23}
\alpha={{n+{1\over2}}\over{\sqrt{12}m}}\cdot\{1+O[(\Omega_{\text{H}}r_{\text{H}})^{-1}]\}\  ,
\end{equation}
which yields the functional expression [see Eq. (\ref{Eq22})]
\begin{equation}\label{Eq24}
r_{\text{min}}=3r_{\text{H}}\Big\{1+{{1}\over{\sqrt{12}m}}\big(n+{1\over2}\big)\cdot
(\Omega_{\text{H}}r_{\text{H}})^{-1}+O[(\Omega_{\text{H}}r_{\text{H}})^{-2}]\Big\}\
\end{equation}
for the radial location of the minimum of the effective binding potential (\ref{Eq16}).

Finally, substituting (\ref{Eq24}) into the relation (\ref{Eq19}), one obtains the discrete
resonance spectrum
\begin{equation}\label{Eq25}
\Omega_0=m\Omega_{\text{H}}\cdot\sqrt{{{32}\over{27}}}
\Big\{1-{{\sqrt{3}}\over{16m}}\big(n+{1\over2}\big)\cdot
(\Omega_{\text{H}}r_{\text{H}})^{-1}+O[(\Omega_{\text{H}}r_{\text{H}})^{-2}]\Big\}\
\end{equation}
which characterizes the 
composed acoustic-black-hole-stationary-bound-state-linearized-massive-scalar-field configurations 
of the photon-fluid model.

\section{Numerical confirmation}

It is of physical interest to test the accuracy of the analytically
derived resonance formula (\ref{Eq25}) which characterizes the 
composed acoustic-black-hole-stationary-scalar-field configurations of the 
photon-fluid model. The corresponding effective field masses $\{\Omega_0(\Omega_{\text{H}},m;n)\}$ of the 
acoustic scalar clouds 
were recently computed numerically in the interesting work \cite{Cim}.

In Table \ref{Table1} we display the dimensionless ratios
$\alpha_{\text{numerical}}$ and $\alpha_{\text{wkb}}$ for the fundamental ($n=0$) resonant 
mode of the stationary bound-state  
acoustic scalar clouds with $m=1$ and for various values of the 
dimensionless angular velocity $\Omega_{\text{H}}r_{\text{H}}$ of the central supporting spinning black hole.
Here $\{\alpha_{\text{numerical}}(\Omega_{\text{H}}r_{\text{H}})\}$ are 
the exact ({\it numerically} computed \cite{Cim}) values of the dimensionless 
ratio $\Omega_0/m\Omega_{\text{H}}$, which characterizes the 
composed acoustic-black-hole-stationary-bound-state-linearized-massive-scalar-field configurations, 
and $\{\alpha_{\text{wkb}}(\Omega_{\text{H}}r_{\text{H}})\}$ are 
the corresponding {\it analytically} derived values of this dimensionless physical parameter as given by the WKB resonance formula (\ref{Eq25}). 

Interestingly, the data presented in Table \ref{Table1} reveals an excellent 
agreement between the numerical data of \cite{Cim} and the
analytically derived WKB resonance formula (\ref{Eq25}) of
the composed acoustic-black-hole-stationary-massive-scalar-field configurations \cite{Notegbb}.

\begin{table}[htbp]
\centering
\begin{tabular}{|c|c|c|}
\hline \ \ $\Omega_{\text{H}}r_{\text{H}}$\ \ & \ \
$\alpha_{\text{numerical}}$\ \ & \ \ $\alpha_{\text{wkb}}$\ \ \\
\hline
\ \ 2\ \ & \ \ 1.058\ \ & \ \ 1.059\ \ \ \\
\ \ 4\ \ & \ \ 1.073\ \ & \ \ 1.074\ \ \ \\
\ \ 6\ \ & \ \ 1.078\ \ & \ \ 1.079\ \ \ \\
\ \ 8\ \ & \ \ 1.081\ \ & \ \ 1.081\ \ \ \\
\hline
\end{tabular}
\caption{Stationary bound-state massive scalar clouds linearly coupled to acoustic spinning black holes 
of the photon-fluid model. We present, for various values of the  
black-hole dimensionless angular velocity $\Omega_{\text{H}}r_{\text{H}}$, 
the exact (numerically computed \cite{Cim}) values of the dimensionless 
ratio $\alpha\equiv\Omega_0/m\Omega_{\text{H}}$ for the fundamental ($n=0$) resonant mode of the stationary spatially regular  
massive scalar clouds with $m=1$. We also present the corresponding analytically derived
values of the dimensionless ratio $\alpha\equiv\Omega_0/m\Omega_{\text{H}}$ as calculated directly from the 
WKB resonance formula (\ref{Eq25}). One finds a remarkably good agreement
between the analytically derived formula (\ref{Eq25}) and the
numerically computed values \cite{Cim} of the dimensionless 
ratio $\Omega_0/m\Omega_{\text{H}}$ which characterizes the composed 
acoustic-black-hole-stationary-bound-state-massive-scalar-field configurations \cite{Notegbb}.}
\label{Table1}
\end{table}

\section{Summary}

The recently published highly important work \cite{Cim} has revealed the physically interesting fact that, in the presence of vortex flows, 
the dynamics of fluctuations in a rotating photon-fluid model is governed by the Klein-Gordon equation of an effective 
massive scalar field in a spinning acoustic black-hole spacetime. 
In particular, it has been demonstrated numerically \cite{Cim} that co-rotating acoustic scalar clouds, 
spatially regular bound-state configurations which are made of stationary linearized massive scalar fields, 
can be supported by the central spinning $(2+1)$-dimensional acoustic black holes. 

The important numerical results presented in \cite{Cim} have nicely demonstrated the fact that, for a given value of the horizon 
angular velocity $\Omega_{\text{H}}$ of the central supporting black hole, the stationary bound-state acoustic clouds of 
the photon-fluid model are characterized by a discrete 
resonance spectrum $\{\Omega_0(\Omega_{\text{H}},m;n)\}^{n=\infty}_{n=0}$ for the effective mass parameter 
of the supported scalar fields. In particular, it has been revealed that the 
composed acoustic-black-hole-stationary-bound-state-massive-scalar-field configurations of the 
photon-fluid model \cite{Cim} exist in the narrow dimensionless 
regime $\alpha\equiv\Omega_0/m\Omega_{\text{H}}\in(1,\alpha_{\text{max}})$ with $\alpha_{\text{max}}\simeq 1.08$.

In the present paper we have used analytical techniques in order to explore the physical and mathematical properties of the
composed bound-state acoustic-black-hole-stationary-linearized-massive-scalar-field configurations. 
In particular, we have derived the remarkably compact WKB analytical formula [see Eq. (\ref{Eq25})]
\begin{equation}\label{Eq26}
{{\Omega_0}\over{m\Omega_{\text{H}}}}=\sqrt{{{32}\over{27}}}\cdot
\Big\{1-{{\sqrt{3}}\over{16m}}\big(n+{1\over2}\big)\cdot
(\Omega_{\text{H}}r_{\text{H}})^{-1}+O[(\Omega_{\text{H}}r_{\text{H}})^{-2}]\Big\}\
\end{equation} 
for the discrete resonant spectrum that 
characterizes the acoustic scalar clouds in the dimensionless 
regime $\Omega_{\text{H}}r_{\text{H}}\gg m$ of rapidly-spinning central supporting black holes. 
The {\it analytically} derived formula (\ref{Eq26}) for the discrete resonant spectrum of the composed 
acoustic-spinning-black-hole-massive-scalar-field configurations was shown to agree remarkably well with direct {\it numerical} computations \cite{Cim} of the corresponding resonant modes of the photon-fluid model. 

Interestingly, from the resonance formula (\ref{Eq26}) one finds the asymptotic upper bound 
\begin{equation}\label{Eq27}
\Big({{\Omega_0}\over{m\Omega_{\text{H}}}}\Big)_{\text{max}}=\sqrt{{{32}\over{27}}}\
\end{equation}
on the regime of existence of the composed acoustic-black-hole-stationary-bound-state-massive-scalar-field 
configurations of the photon-fluid model. Our results therefore provide a simple analytical explanation for the intriguing upper 
bound (\ref{Eq3}) on the regime of existence of the cloudy acoustic black-hole spacetimes 
that has recently been observed numerically in the physically interesting work \cite{Cim}.  

\bigskip
\noindent
{\bf ACKNOWLEDGMENTS}
\bigskip

This research is supported by the Carmel Science Foundation. I thank
Yael Oren, Arbel M. Ongo, Ayelet B. Lata, and Alona B. Tea for
stimulating discussions.



\begin{thebibliography}{99}

\bibitem{Kerr} R. P. Kerr, Phys. Rev. Lett. {\bf 11}, 237 (1963).

\bibitem{Chan} S. Chandrasekhar, {\it The Mathematical Theory of Black
Holes}, (Oxford University Press, New York, 1983).

\bibitem{Hodrc} S. Hod, Phys. Rev. D {\bf 86}, 104026 (2012) [arXiv:1211.3202];
S. Hod, The Euro. Phys. Journal C {\bf 73}, 2378 (2013)
[arXiv:1311.5298]; S. Hod, Phys. Rev. D {\bf 90}, 024051 (2014)
[arXiv:1406.1179]; S. Hod, Phys. Lett. B {\bf 739}, 196 (2014)
[arXiv:1411.2609]; S. Hod, Class. and Quant. Grav. {\bf 32}, 134002
(2015) [arXiv:1607.00003]; S. Hod, Phys. Lett. B {\bf 751}, 177
(2015); S. Hod, Class. and Quant. Grav. {\bf 33}, 114001 (2016); S.
Hod, Phys. Lett. B {\bf 758}, 181 (2016) [arXiv:1606.02306]; S. Hod
and O. Hod, Phys. Rev. D {\bf 81}, 061502 Rapid communication (2010)
[arXiv:0910.0734]; S. Hod, Phys. Lett. B {\bf 708}, 320 (2012)
[arXiv:1205.1872]; S. Hod, Jour. of High Energy Phys. {\bf 01}, 030
(2017) [arXiv:1612.00014].

\bibitem{HerR} C. A. R. Herdeiro and E. Radu, Phys. Rev. Lett. {\bf 112}, 221101
(2014); C. L. Benone, L. C. B. Crispino, C. Herdeiro, and E. Radu,
Phys. Rev. D {\bf 90}, 104024 (2014); C. A. R. Herdeiro and E. Radu,
Phys. Rev. D {\bf 89}, 124018 (2014); C. A. R. Herdeiro and E. Radu,
Int. J. Mod. Phys. D {\bf 23}, 1442014 (2014); Y. Brihaye, C.
Herdeiro, and E. Radu, Phys. Lett. B {\bf 739}, 1 (2014); J. C.
Degollado and C. A. R. Herdeiro, Phys. Rev. D {\bf 90}, 065019
(2014); C. Herdeiro, E. Radu, and H. R\'unarsson, Phys. Lett. B {\bf
739}, 302 (2014); C. Herdeiro and E. Radu, Class. Quantum Grav. {\bf
32} 144001 (2015); C. A. R. Herdeiro and E. Radu, Int. J. Mod. Phys.
D {\bf 24}, 1542014 (2015); C. A. R. Herdeiro and E. Radu, Int. J.
Mod. Phys. D {\bf 24}, 1544022 (2015); P. V. P. Cunha, C. A. R.
Herdeiro, E. Radu, and H. F. R\'unarsson, Phys. Rev. Lett. {\bf
115}, 211102 (2015); B. Kleihaus, J. Kunz, and S. Yazadjiev, Phys.
Lett. B {\bf 744}, 406 (2015); C. A. R. Herdeiro, E. Radu, and H. F.
R\'unarsson, Phys. Rev. D {\bf 92}, 084059 (2015); C. Herdeiro, J.
Kunz, E. Radu, and B. Subagyo, Phys. Lett. B {\bf 748}, 30 (2015);
C. A. R. Herdeiro, E. Radu, and H. F. R\'unarsson, Class. Quant.
Grav. {\bf 33}, 154001 (2016); C. A. R. Herdeiro, E. Radu, and H. F.
R\'unarsson, Int. J. Mod. Phys. D {\bf 25}, 1641014 (2016); Y.
Brihaye, C. Herdeiro, and E. Radu, Phys. Lett. B {\bf 760}, 279
(2016); Y. Ni, M. Zhou, A. C. Avendano, C. Bambi, C. A. R. Herdeiro,
and E. Radu, JCAP {\bf 1607}, 049 (2016); M. Wang, arXiv:1606.00811
.

\bibitem{Zel} Ya. B. Zel`dovich, Pis`ma Zh. Eksp. Teor. Fiz. {\bf
14}, 270 (1971) [JETP Lett. {\bf 14}, 180 (1971)]; Zh. Eksp. Teor.
Fiz. {\bf 62}, 2076 (1972) [Sov. Phys. JETP {\bf 35}, 1085 (1972)];
A. V. Vilenkin, Phys. Lett. B {\bf 78}, 301 (1978).

\bibitem{PressTeusup} W. H. Press and S. A. Teukolsky, Nature {\bf 238}, 211 (1972); 
W. H. Press and S. A. Teukolsky, Astrophys. J.
{\bf 185}, 649 (1973).

\bibitem{Noteunits} We use natural units in which $G=c=\hbar=1$.

\bibitem{Notedim} Note that the mass parameter $\mu$ of the supported scalar field 
stands for $\mu/\hbar$ and therefore, like the frequency $\omega$, has the dimensions of (length)$^{-1}$.

\bibitem{Cim} M. Ciszak and F. Marino, arXiv:2101.07508 .

\bibitem{PF1} T. Frisch, Y. Pomeau, S. Rica, Phys. Rev. Lett. {\bf 69}, 1644 (1992).

\bibitem{PF2} Y. Pomeau, S. Rica, Phys. Rev. Lett. {\bf 71}, 247 (1993).

\bibitem{Un} W. G. Unruh, Phys. Rev. Lett. {\bf 46}, 1351 (1981).





\bibitem{Mar1} F. Marino, Phys. Rev. A {\bf 100}, 063825 (2019).

\bibitem{Notemmm} Here $m$ is the azimuthal harmonic index of the stationary acoustic scalar mode [see Eq. (\ref{Eq6}) below]. 

\bibitem{Notennnn} The parameter $n$, which characterizes the discrete resonance spectrum of the 
composed acoustic-black-hole-stationary-scalar-clouds configurations, is an integer [see Eq. (\ref{Eq17}) below].

\bibitem{Noterhrh} Here $r_{\text{H}}$ is the horizon radius of the central supporting acoustic black hole.

\bibitem{Notert} Here the coordinates $\{r,\theta\}$ are respectively the radial and azimuthal polar
coordinates in the $xy$ plane. 


\bibitem{Notegrr} Note that $g_{rr}\to\infty$ at the horizon $r=r_{\text{H}}$ of the effective acoustic spacetime (\ref{Eq4}). 

\bibitem{Notecs} We shall use natural units in which $c_{\text{s}}\equiv1$.

\bibitem{Notem11} We shall henceforth assume $m>0$ for the stationary spatially regular acoustic scalar modes. 

\bibitem{Notenub} It is interesting to note that, unlike spinning Kerr black holes whose 
horizon angular velocity is bounded from above by the simple dimensionless relation $\Omega_{\text{H}}r_{\text{H}}\leq 1/2$, the acoustic black-hole spacetimes of the photon-fluid model have no strict upper bound on their angular 
velocities \cite{Cim}. 

\bibitem{Notemap} Note that the semi-infinite radial range $r\in [r_{\text{H}},\infty]$ of the acoustic black-hole spacetime 
is mapped into the infinite 
range $y\in [-\infty,+\infty]$ by the radial differential relation (\ref{Eq15}).

\bibitem{WKB1} B. F. Schutz and C. M. Will, Astrophys. J. {\bf 291}, L33 (1985).

\bibitem{WKB2} S. Iyer and C. M. Will, Phys. Rev. D {\bf 35}, 3621 (1987).

\bibitem{WKB3} S. Iyer, Phys. Rev. D {\bf 35}, 3632 (1987).

\bibitem{Will} L. E. Simone and C. M. Will, Class. Quant. Grav. {\bf 9}, 963 (1992).

\bibitem{Hodalm} S. Hod, Phys. Lett. B {\bf 746}, 365 (2015) [arXiv:1506.04148].

\bibitem{Notegbb} It is worth noting that the agreement between the numerical data of \cite{Cim} and the
analytically derived WKB resonance formula (\ref{Eq25}) of the 
composed acoustic-black-hole-stationary-bound-state-linearized-massive-scalar-field configurations 
is generally better than $0.1\%$ in the $\Omega_{\text{H}}r_{\text{H}}\gtrsim1$ regime. 
This observation is quite remarkable 
since the analytically derived WKB resonance spectrum (\ref{Eq25}) is formally valid 
in the eikonal large-$\Omega_{\text{H}}r_{\text{H}}$ regime. 

\end{thebibliography}
\end{document}